\documentclass[preprint,showpacs,preprintnumbers,amsmath,amssymb,superscriptaddress]{revtex4}

\usepackage{graphicx}
\usepackage{dcolumn}
\usepackage{bm}

\begin{document}


\title{Thermal shot noise in top-gated single carbon nanotube field effect transistors}

\author{J. Chaste}
\affiliation{Laboratoire Pierre Aigrain, ENS-CNRS-UMR8551,
24 rue Lhomond, 75005 Paris, France }
\author{E. Pallecchi}
\affiliation{Laboratoire Pierre Aigrain, ENS-CNRS-UMR8551,
24 rue Lhomond, 75005 Paris, France }
\author{P. Morfin}
\affiliation{Laboratoire Pierre Aigrain, ENS-CNRS-UMR8551,
24 rue Lhomond, 75005 Paris, France }
\author{G. F\`eve}
\affiliation{Laboratoire Pierre Aigrain, ENS-CNRS-UMR8551,
24 rue Lhomond, 75005 Paris, France }
\author{T. Kontos}
\affiliation{Laboratoire Pierre Aigrain, ENS-CNRS-UMR8551,
24 rue Lhomond, 75005 Paris, France }
\author{J.-M. Berroir}
\affiliation{Laboratoire Pierre Aigrain, ENS-CNRS-UMR8551,
24 rue Lhomond, 75005 Paris, France }
\author{P. Hakonen}
\affiliation{
Low Temperature Laboratory, Aalto University,
Puumiehenkuja 2 B, FI-00076 Aalto, Finland}
\author{B. Pla\c{c}ais}
\email{Bernard.Placais@lpa.ens.fr}
\affiliation{Laboratoire Pierre Aigrain, ENS-CNRS-UMR8551,
24 rue Lhomond, 75005 Paris, France }

\date{\today}

\begin{abstract}

The high-frequency transconductance and current noise of top-gated single carbon nanotube transistors have been measured and used to investigate hot electron effects in one-dimensional transistors. Results are in good agreement with a theory of 1-dimensional nano-transistor. In particular the prediction of a large transconductance correction to the Johnson-Nyquist thermal noise formula is confirmed experimentally. Experiment shows that nanotube transistors can be used as fast charge detectors for quantum coherent electronics with a resolution of $13\mathrm{\mu e/\sqrt{Hz}}$ in the $0.2$--$0.8\;\mathrm{GHz}$ band.

\end{abstract}

\maketitle


Field effect transistors (FETs), such as quantum point contact transistors \cite{Gustavsson2008NL}, are an alternative to Coulomb blockade devices (SETs)\cite{Andresen2008JAP} for fast single charge detection in nanostructures due to their smaller impedance. In the perspective of a future quantum electronics based on ballistic electrons \cite{Feve2007Science}, or the fast readout of charge qubits,  a nanosecond time resolution is needed that is challenging but should be reachable by using nano-transistors. A well-known realization is a top gated single carbon nanotube  \cite{Tans1998Nature}, which works at high frequency \cite{Chaste2008NanoL} (see review in \cite{Rutherghen2009NatMat}). Ultimate gate coupling and finite
density of states in the channel push these nano-transistors  close to the quantum limit where gate capacitance $C_g$  approaches the quantum capacitance $C_q$ \cite{Chaste2008NanoL}. Beside large charge sensitivity, an important limiting factor of resolution is thermal noise from hot electrons which is prominent
in low-dimensional conductors due to poor energy relaxation \cite{Pothier1997PRL,Lazzeri2006PRB}. Effect of dissipation can be investigated by measuring the out-of-equilibrium  phonon population at finite bias $V_d$ \cite{Steiner2009NatureNanotech} but also by noise thermometry \cite{Wu2007PRB}. The nano-transistor geometry
offers an opportunity to revisit these hot electron effects using the additional control of electronic transmission.

The noise thermometry approach of electronic population  relies on assumption of a thermal distribution
and the Johnson-Nyquist formula $S_I=4g_nk_BT_e$   relates the current noise spectral density $S_I$  to the electronic temperature $T_e$ via a
 so-called "noise conductance"  $g_n$ \cite{van_der_Ziel1962IEEE}.
In general one has  $g_n=g_d$, the differential drain conductance. The situation is different in
gated semiconductors where an additional contribution arises, at finite bias, associated with transconductance $g_m=\partial I_d/\partial V_g$  where $I_d$ is the drain current
and $V_g$ the  gate voltage.  This term depends \textit{a priori} on geometry, at least for the 3D and 2D transistors
\cite{van_der_Ziel1962IEEE,Naveh1999PRB}.  We show here that the situation is different at 1D and that a simple relation exists between
$g_d$, $g_m$ and $g_n$ which only brings in the gate coupling factor $\beta=C_g/C_q$.

The paper reports on current noise and transconductance of top-gated single nanotube transistors.
Measurements were carried at 4K to take advantage of the enhanced noise resolution of cryogenic setups and in a GHz bandwidth to overcome the  $1/f^\alpha$ low frequency noise and benefit from the good AC coupling in the contact impedance.
Hot electron effects show up both in the thermal current noise and the width of the transconductance peak at the onset
of conduction.
Using a 1D nano-transistor model based on scattering theory \cite{Blanter2000PhysRep} we  obtain a generalized Johnson-Nyquist noise formula in the form,
\begin{equation}
S_I= 4\left(g_{d}+g_m/2\beta\right) k_BT_e\label{S_I}\; ,
\end{equation}
where  conductance terms and noise are to be taken at the same frequency.  Eq.(\ref{S_I}) is supported by the RF measurements of  transconductance and noise and the electronic temperature reported below as function of bias voltage.
As electronic temperature depends on current, one can alternatively express thermal noise as  $S_I=2eI_d\tilde F$,  by introducing a pseudo Fano factor $\tilde F\lesssim1$. The limit $\tilde F=1$ corresponds to a classical shot noise as observed in vacuum diodes. The hot electron regime shows up in our nanotube transistor by a full thermal shot noise with  $\tilde F\sim1$ at low bias followed by some reduction  ($\tilde F\simeq0.7$) resulting from Pauli principle and the effect of electronic degeneracy which generally shows up at high bias.

\begin{figure}
\includegraphics[scale=0.8]{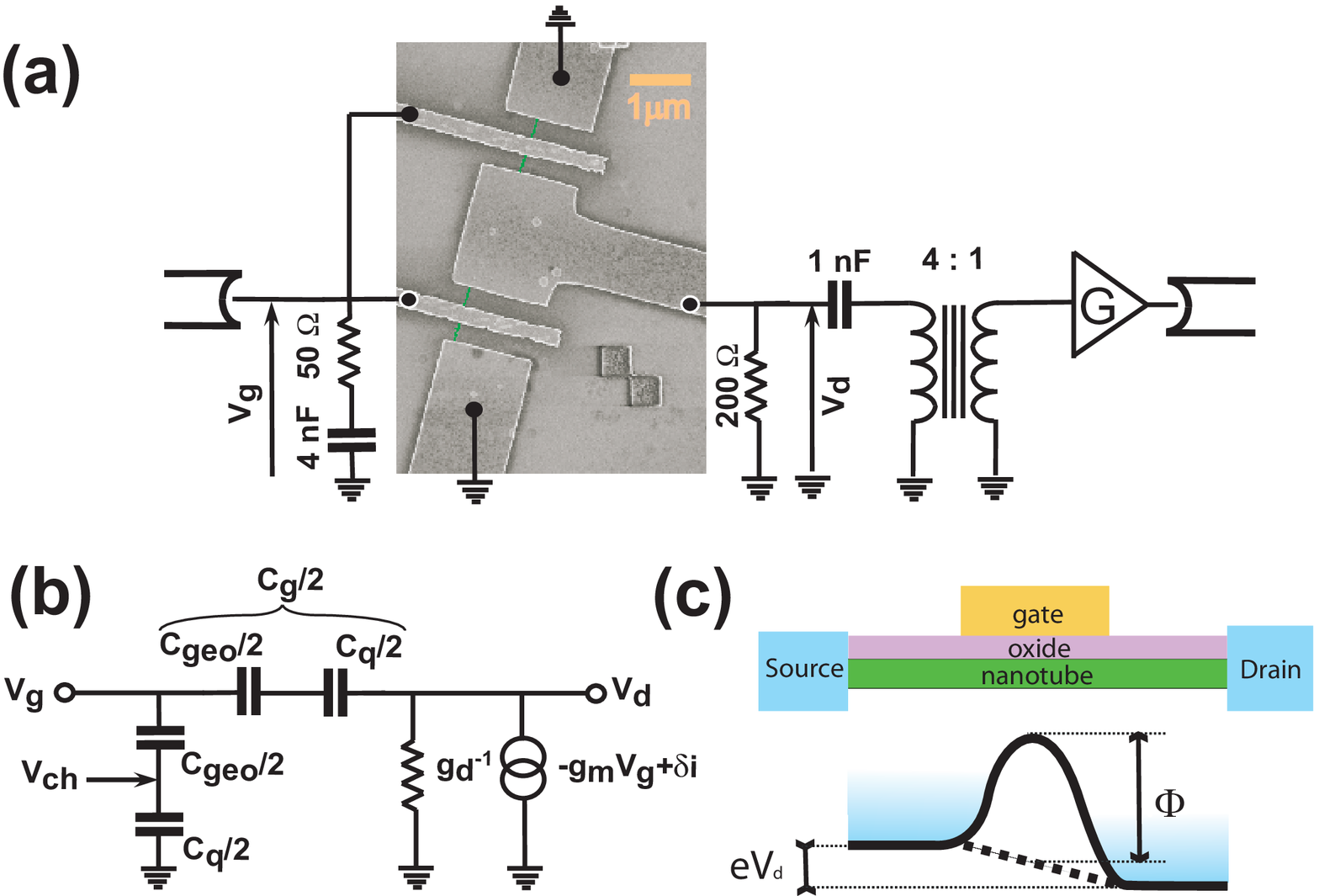}
\caption{(a) A double gate nanotube transistor embedded in the (4K) measuring scheme which includes bias resistors,  DC blocks and a cryogenic low noise amplifier with a $200:50\;\mathrm{Ohms}$ matching transformer. The nanotube (colored in green), of diameter $\sim 2\;\mathrm{nm}$, is covered in its middle part by a top gate (length $300\;\mathrm{nm}$) which controls the local barrier. The leads on both sides act as electronic reservoirs.   (b) The lumped element description of the nanotube transistor RF response includes the channel resistance $g_d^{-1}$, the transconductance $g_m$,
the gate capacitance $C_g=C_ {geo}C_q/(C_ {geo}+C_q)$ with its geometrical (resp. quantum) contributions $C_{geo}$ (resp. $C_q$) and the noise current generator $\delta i$.
The channel potential $V_{ch}=V_g\times C_g/C_q$
governs the barrier height $\Phi=-e(V_{ch}+Const.)$ which defines  the gate coupling coefficient $\beta=\partial\Phi/\partial V_g=C_g/C_q$. (c) Sketch of the energy profile (solid line) and the electronic distribution (blue gray scale) of a top-gated nanotube transistor. }
 \label{schema}
\end{figure}

The sample (Fig.\ref{schema}(a)) is taken from a batch which was  extensively described and characterized in Ref.\cite{Chaste2008NanoL}.
A symmetric double gate RF design is used (Fig.\ref{schema}(a)) on high resistivity silicon substrate. The high mobility CVD-grown nanotube (diameter
$d\sim2\;\mathrm{nm}$) is equipped with a top gate (length $L_g=0.3\mathrm{\mu m}$) deposited on a thin AlO$_x$ oxide (thickness $t_{ox}\simeq6\;\mathrm{nm}$). Palladium drain and source metallisations
are used for low Schottky barrier contacts. Our high-sensitivity cryogenic setup includes a low noise amplifier fitted to a $200$--$50\;\mathrm{Ohms}$ impedance matching transformer with a $0.8\;\mathrm{GHz}$ cutoff.
Matched resistors are fitted at the input and output lines to obtain a broad $0.1$--$0.8\;\mathrm{GHz}$ measuring band.
The  200 Ohms output load ensures DC voltage bias conditions and serves as an auxiliary white noise source
for in-situ calibration. The lumped circuit element description of the nanotube transistor (Fig.\ref{schema}(b)) is used for
RF data analysis and the theoretical model below.
The gate capacitance, $C_g/L_g\simeq0.07\pm0.02\;\mathrm{fF/\mu m}$, is taken from the room temperature  RF probe station measurements \cite{Chaste2008NanoL}. With $C_q/L_g=8e^2/hv_F=0.4\;\mathrm{fF/\mu m}$ ($v_F\simeq8\times 10^5\;\mathrm{m/s}$)
we deduce $\beta\simeq0.2$  \cite{Chaste2008NanoL}. We have used negative drain bias, which shows lower $1/f^\alpha$ noise and
symmetric gate bias conditions.
$I_d$ and $S_I$ are taken by reference to the
pinch-off value (at $V_g=+1\;\mathrm{V}$).

\begin{figure}
\includegraphics[scale=0.8]{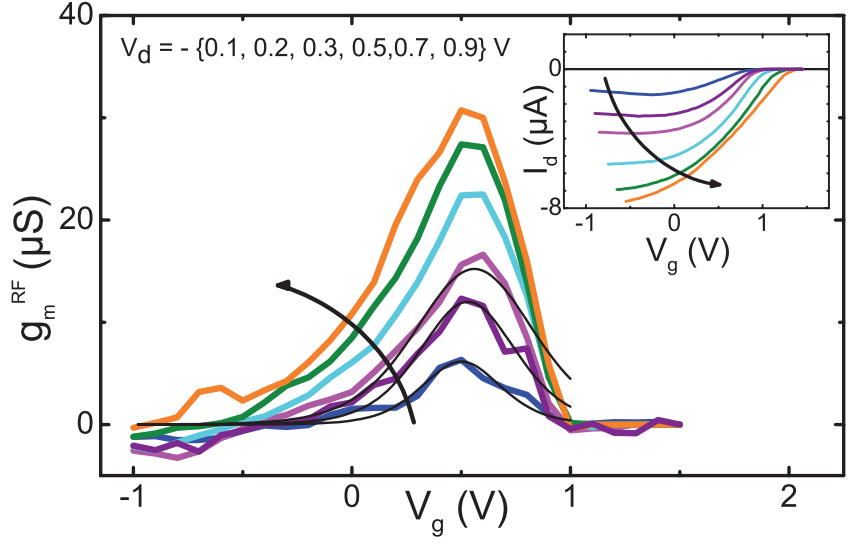}
\caption{Carbon nanotube transistor characteristics including the DC
current (inset) and the  RF transconductance $g_m^{RF}$ measured in the $0-0.8\;\mathrm{GHz}$ band (main panel) as function of gate voltage $V_g$ for a representative set of (color encoded) bias voltages increasing according to the arrow.
The transconductance maximum for $V_g\simeq+0.5\;\mathrm{V}$  saturates at a large value $\gtrsim30\;\mathrm{\mu S}$.
Low bias  data, for $V_{d}=-(100,200,300)\;\mathrm{mV}$, are fitted with Eq.(\ref{g_m}) (solid lines) within a prefactor $\simeq0.6$  added to account for non ideal ballistic transport above the barrier.
The fitting parameter, $V_{Te}=k_BT_e/e\beta\simeq(120,140,170)\pm0.01\;\mathrm{mV}$,  gives (with $\beta=0.2$) the thermal energies
$k_BT_e=(24,28,34)\;\mathrm{meV}$.} \label{ad128_dc_et_RF}
\end{figure}

Figure \ref{ad128_dc_et_RF} shows the radio frequency transconductance $g_m^{RF}$ deduced from transmission measurements
\cite{Chaste2008NanoL} as function of gate voltage for different bias conditions. The DC conductance
 $g_d^{DC}$ and  transconductance $g_m^{DC}$ are obtained from the $I_d(V_g,V_d)$ characteristics (Fig.\ref{ad128_dc_et_RF}-inset).
Reflection coefficients, and $g_d^{RF}$, cannot be accessed with this setup.
The sample shows large $g_m^{RF}\simeq30\;\mathrm{\mu S}$, typically 3-times larger than $g_m^{DC}$, which suggests AC contact coupling. We observe a small temperature dependence of $g_m^{RF}$ (a $\sim 50\%$ decrease between 4K and 300K) which we take as a first indication of a hot electron regime.

\begin{figure}
\includegraphics[scale=0.8]{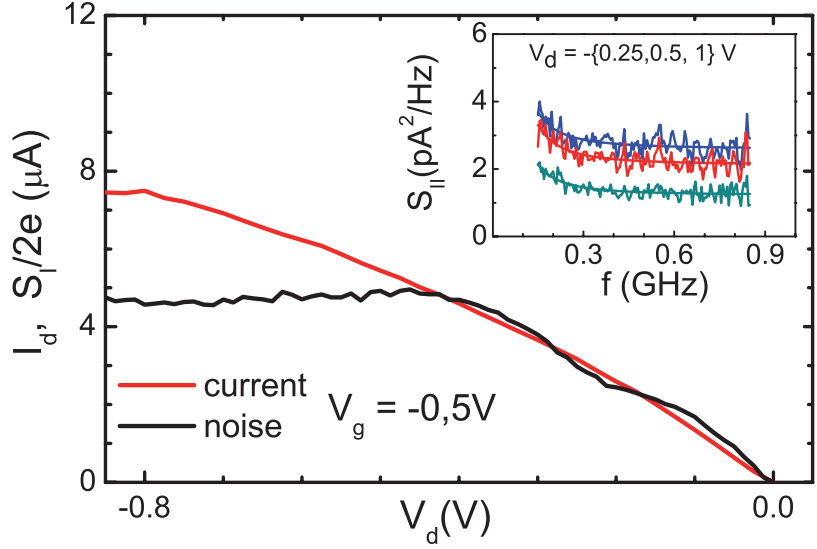}
\caption{
Bias dependence of the DC current and current noise of the nanotube transistor in the metallic state ($V_g\leq0$). The intrinsic nanotube shot noise $S_I$ is deduced by fitting the raw noise spectra (inset) with a  $A+B/f^2$ law (solid lines) and subtracting a ($B/f^2$) Lorenzian tail for the environment noise which contributes below $0.5\;\mathrm{GHz}$ and the white noise ($S_I/V_d\simeq1.1\;\mathrm{(pA)^2V^{-1}}$) of the $200\;\mathrm{Ohm}$ bias resistance, which is independently measured at the pinch-off.  Accurate shot noise data, in the $0.1$--$0.8\;\mathrm{GHz}$ band, are plotted in the main panel as function of bias voltage for comparison with DC current. We observe full shot-noise ($S_I/2e\simeq I_d$) for $V_d>-0.4\;\mathrm{V}$ and a noise saturation for $V_d<-0.4\;\mathrm{V}$. } \label{metallic}
\end{figure}

In Fig.\ref{metallic} we discuss measurements of the transistor in the open state ($V_g=-0.5\;\mathrm{V}$) where $g_m^{RF}\simeq0$.
Here  the nanotube behaves as a metallic wire with a finite transmission $D\sim0.1$, deduced from low bias current and presumably limited by contact barriers.
 As seen in  the figure, a  full shot noise limit is observed for  $|V_d|\lesssim 0.4\;\mathrm{V}$,  followed by a saturation plateau. Although the noise characterization is certainly
 relevant for understanding the saturation mechanism that takes place in nanotubes at very high bias, we prefer to leave this discussion for a future work.
Using $S_I=2eI_d{\tilde F}\gtrsim 4g_d^{DC}k_BT_e$ ($g_m\simeq0$ and $g_d\gtrsim g_d^{DC}$ in Eq.(\ref{S_I})) we deduce ${\tilde F}\gtrsim 2k_BT_e/eV_d$ and an (over-) estimate of electronic temperature $k_BT_e/eV_d\lesssim {\tilde F}/2\simeq0.6$. Indeed, this ratio is larger than a theoretical expectation,
 $k_BT_e/eV_d\sim \sqrt{6D/\pi^2} \sim 0.25$, for a Wiedemann-Franz resistance limited thermal sink \cite{Kumar1996PRL}. Importantly, both numbers show that a hot electron regime is expected and present in the nanotube. As discussed below, we can rely on the transconductance characteristics for an independent determination of the electronic temperature of the nanotube working as a transistor.

\begin{figure}
\includegraphics[scale=1]{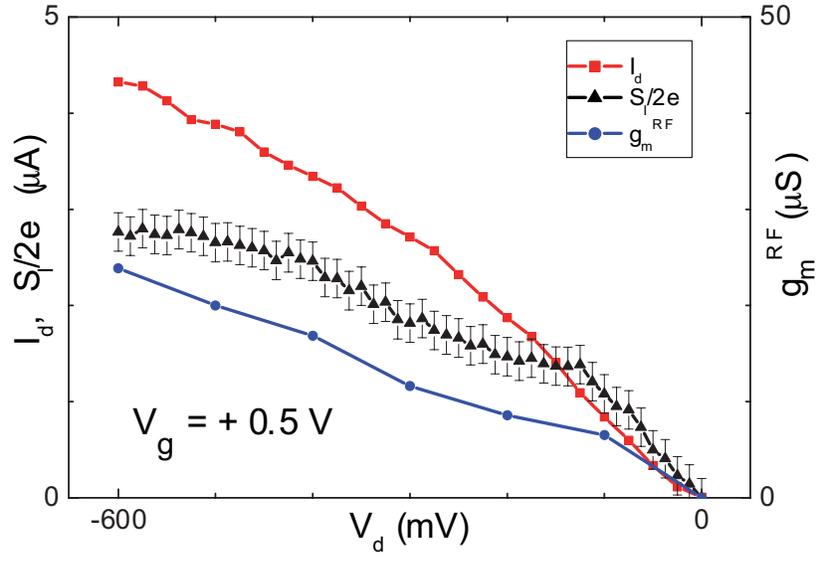}
\caption{Bias dependence of DC current $I_d$ (squares), RF current noise $S_I$ (triangles) and RF transconductance $g_m$ (circles) at the transconductance maximum ($V_g\sim+0.5\;\mathrm{V}$). } \label{Noise temp}
\end{figure}

 Fig.\ref{Noise temp} contains the main experimental results of the paper, namely the bias dependence of  $I_{d}$, $S_I$ and $g_m$,
 at the transconductance maximum ($V_g=+0.5\;\mathrm{V}$). The $S_I(V_d)$ dependence is similar to that observed in the open state with, as a difference,  a first noise plateau ($\tilde F\simeq0.7$) observed in range  $|V_d|=0.1$--$0.3\;\mathrm{V}$.
The two step increase of noise and the width of the plateau ($\sim0.3\;\mathrm{V}$) are suggestive of the subband structure of the nanotube (intersubband-gap $\sim0.3$--$0.5\;\mathrm{eV}$ for a $2\;\mathrm{nm}$ diameter).  In order to secure 1D transport conditions, we shall therefore focus below to the ($|V_d|\lesssim0.3\;\mathrm{V}$) bias range.

For the theoretical analysis we rely on a simple 1D nano-transistor model.
The nanotube is described as a fourfold degenerate 1D channel decomposed in three sections (see Fig.\ref{schema}(c)):
a central part covered by the top-gate which acts as a local classical barrier and two ungated nanotube leads on both sides which constitute the drain and source reservoirs of the transistor. The leads are diffusive and assumed to be populated with Fermi distributions $f_s(E)$ and $f_d(E)=f_s(E+eV_{d})$. The
barrier  (height $\Phi$) acts as an high-pass energy filter with quasi-ballistic high energy electrons (transmission $D(E)\lesssim1$ for $E>\Phi$) and reflected
low energy ones ($D(E)=0$ for $E<\Phi$). With these assumptions and scattering theory  \cite{Blanter2000PhysRep} one can readily calculate current and noise to deduce the differential
conductance $g_{d}= \frac{4e^2}{h} f_d(\Phi)$,
the transconductance
\begin {equation}
g_{m}= \beta\frac{4e^2}{h}[f_s(\Phi)-f_d(\Phi)]\qquad,\label{g_m}
\end{equation}
and the noise conductance  $g_{n}= \frac{2e^2}{h}[f_s(\Phi)+f_d(\Phi)]$. According to the equivalent circuit in Fig.\ref{schema}(b), we have taken $\partial\Phi/\partial  V_g=-e\beta$.  These three "conductances" depend on two parameters,  the
drain and source occupation numbers (at $E=\Phi$), and therefore obey the constitutive relation
$g_n=g_{d}+g_m/2\beta$ which gives rise to Eq.(\ref{S_I}). Details of screening,
which are encoded in the $\Phi(V_d,V_g)$ functional, or temperature $T_e(V_d,V_g)$ in $f_{s,d}(\Phi)$, are implicit in  Eq.(\ref{S_I}) which can  be regarded as a universal result for 1D transistors.
In terms of thermal shot-noise one has   ${\tilde F}=\coth{(eV_{d}/2k_BT_e)}\lesssim1.3$ (for $2k_BT_e/eV_d\lesssim 1$) at low bias where $g_m<2\beta g_d$.
At high bias one has $g_m\gg 2\beta g_d$  whenever  $f_d(\Phi)\ll f_s(\Phi)\lesssim1$; one also expect Pauli noise suppression with a reduction factor $(1-f_s(\Phi))$ like in the usual 2D case \cite{Naveh1999PRB}.

As a first test of the model we obtain a good fit of the $g_m^{RF}(V_g)$ data in Fig.\ref{ad128_dc_et_RF} with Eq.(\ref{g_m}) by taking $\Phi(V_g)=Const.-e\beta V_g$ and $E_F\ll \Phi$ in $f_{s,d}(\Phi)$ according to the equivalent circuit  and  the potential landscape in Figs.\ref{schema}(b) and (c). We have used a scaling factor $0.6$ that accounts for residual electronic diffusion above the barrier. Electronic temperatures deduced from theoretical fits   corresponds to a large absolute temperature ($k_BT_e\simeq28\;\mathrm{meV}$ at $V_d\sim200\;\mathrm{mV}$), but still a small energy spread when compared to the bias voltage ($k_BT_e\simeq0.14\; eV_d$) .

The model accounts qualitatively for a full shot-noise at low bias (${\tilde F}\simeq1$ for $V_d\lesssim0.1\;\mathrm{V}$) and for the noise reduction at intermediate bias ($|V_d|=0.1$--$0.3\;\mathrm{V}$).
 Using the electronic temperature from transconductance fits in Fig.\ref{ad128_dc_et_RF}, we can \emph{quantitatively}  compare noise and transconductance in Eq.(\ref{S_I}):
 At $V_{d}=-0.2\;\mathrm{V}$, where $S_I/2e\simeq1.46\pm0.1\;\mathrm{\mu A}$, $g_m^{RF}\simeq8.6\pm0.5\;\mathrm{\mu S}$ and
 $V_{Te}=k_BT_e/e\beta\simeq0.14\pm0.02\;\mathrm{V}$ we obtain $g_m^{RF}V_{Te}\simeq1.2\pm0.3\;\mathrm{\mu A}$
in good agreement with  theoretical prediction $S_I/2e\gtrsim g_mV_{Te}$ from Eq.(\ref{S_I}).
By comparison  standard noise thermometry would give a smaller value
$S_I/2e= 2g_dk_BT_e/e\simeq 0.5 \;\mathrm{\mu A}$ (taking $g_d=g_d^{DC}\simeq9\;\mathrm{\mu S}$). We take this agreement as a strong support for the 1D model.
Further experimental investigations should involve direct measurement of $g_d^{RF}$ to test Eq.(\ref{S_I}) over the full bias
range including the crossover  $g_m^{RF}\lesssim 2\beta g_d^{RF}$.

Finally we can use our data to estimate the charge resolution
$\delta q_{rms}=\sqrt{S_I}C_g/g_m$.
The best signal to noise conditions, for  $V_d\sim0.3\;\mathrm{V}$  and
  $I_dV_d\sim1\;\mathrm{\mu W}$ where $g_m\gtrsim15\;\mathrm{\mu S}$
 and $S_I/2e \lesssim2\;\mathrm{\mu A}$, give
 $\delta q_{rms}\lesssim 13\mathrm{\mu e/\sqrt{Hz}}$ ($C_g\lesssim 40\;\mathrm{aF}$ \cite{Chaste2008NanoL}) for our double gate device
which corresponds to an \emph{rms} charge resolution of 0.4 electron in the  $0.8\;\mathrm{GHz}$ bandwidth of our set-up.
Within a factor five of the best resolution achieved in SETs \cite{Andresen2008JAP}, this
smaller sensitivity of the present CNT-FETs  is
balanced by a much larger bandwidth ($0.8\;\mathrm{GHz}$ here against $0.08\;\mathrm{GHz}$ in \cite{Andresen2008JAP}) and the possibility to operate at room temperature.

In conclusion, our comprehensive study of high-frequency transport and shot noise has confirmed that the single nanotube transistor
is a model system and that hot electron effects are prominent at 1D. In particular our data support a generalized Johnson-Nyquist expression for thermal noise in 1D transistors introduced in the paper.
Finally we have benchmarked nanotube FETs against nanotube SETs for applications as fast single electron detectors.

\begin{acknowledgments}
Authors acknowledge fruitful discussions with  G. Dambrine, V. Derycke, P. Dollfus, C. Glattli, H. Happy and M. Sanquer. The research has been supported by french ANR under contracts ANR-2005-055-HF-CNT, ANR-05-NANO-010-01-NL-SBPC, and EU-STREP
project CARDEQ under contract IST-FP6-011285.
\end{acknowledgments}

\newpage

\end{document}